\documentclass[prl,twocolumn,showpacs,preprintnumbers,amsmath,amssymb,superscriptaddress]{revtex4}

\usepackage{graphicx}
\usepackage{dcolumn}
\usepackage{bm}

\newcommand{\ket}[1]{\ensuremath| #1 \rangle}

\begin{document}

\title{Direct, Loss-Tolerant Characterization of Nonclassical Photon Statistics}

\author{Daryl Achilles}
\email{daryl.achilles@physics.ox.ac.uk}
  \affiliation{Clarendon Laboratory, University of Oxford, Parks Road, Oxford, OX1 3PU,United Kingdom}

\author{Christine Silberhorn}
  \affiliation{Clarendon Laboratory, University of Oxford, Parks Road, Oxford, OX1 3PU,United Kingdom}
     \affiliation{Max Planck Research Group, G\"unther Scharowskystr. 1 / Bau 24, 91058 Erlangen, Germany }

\author{Ian A. Walmsley}
  \affiliation{Clarendon Laboratory, University of Oxford, Parks Road, Oxford, OX1 3PU,United Kingdom}

\begin{abstract}
We  experimentally investigate a method of directly characterizing
the photon number distribution of nonclassical light beams that is
tolerant to losses and makes use only of standard binary detectors.
This is achieved in a single measurement by calibrating the detector
using some small amount of prior information about the source. We
demonstrate the technique on a freely propagating heralded two-photon number state
 created by conditional detection of a two-mode squeezed
state generated by a parametric downconverter.
\end{abstract}


\date{\today}

 \maketitle



The photon number distribution is a key characteristic of every
optical field.  Several indicators of nonclassicality are based
directly on the measurement of these statistics, including negativity of Mandel's
$Q$ Parameter and the negativity of the Glauber-Sudarshan $P$
function~\cite{Mandel}.  Furthermore many quantum information
applications rely on light sources with well-defined photon number
distributions. For example, quantum cryptography based on
single-photon protocols~\cite{QKD} requires the complete suppression
of multi-photon components in order to guarantee security over
longer distances. The ability to directly measure the photon number
distributions, then, is important both fundamentally and
technologically.  Two main obstacles have hindered progress in such
measurements: first, photon-number-resolving detectors were only
devised recently \cite{Loopy, Yamamoto_NIST}; second, all
available single-photon sensitive detectors exhibit finite
efficiencies, such that intrinsic losses often mask the signatures
of nonclassical states.

Currently there are three approaches to retrieving photon number
distributions: using photon-number-resolving
detectors~\cite{Yamamoto_NIST}; via state reconstruction from
homodyne tomography~\cite{Tomo, Mlynek}; and
using binary (``click-counting'') detectors like avalanche
photodiodes (APDs)~\cite{Loopy, Attenuation, Paris}.
However, all of these approaches are compromised by loss and
detector inefficiencies, which cause instabilities in the inversion
algorithm used to reconstruct photon number distributions from count
statistics. This makes it difficult to reconstruct the quantum state
of the source since the detector efficiency must be known accurately. Several photon-number-resolving detectors have high
detector efficiencies, though these are usually accompanied by noise
from dark counts, which also affects the fidelity of the inversion.
In homodyne detection the calibration is made yet more difficult by
the need to match the modes of the quantum state with that of the
local oscillator.  Two distinct approaches are currently known using
APDs: measurements of the mean count rate as a function of beam
attenuation~\cite{Attenuation, Paris} and
mode-multiplexing to implement a photon-number-resolving
detector~\cite{Loopy}. Photon number characteristics for classical
states have been measured with APDs in the past~\cite{Loopy,
Loop, Paris}, but the problem is much more intricate
for nonclassical states, which do not maintain the form of their
statistics when attenuated.  Absolute APD quantum efficiency
measurements usually require an independent measurement in which the
detector response can be distinguished from external losses.  Poor
calibration may compromise the accuracy of methods such as the
attenuation approach that rely on well-known losses.  In contrast,
the mode-multiplexing approach greatly increases the accuracy of
detector calibration and does not require an independent
measurement.



In this Letter we propose and demonstrate a new APD-based approach
to direct loss-tolerant photonic state characterization. We exploit
partial {\em a priori} information about the photon number
distribution, typically known from the state generation process, to
accurately calibrate the total loss in the channel taken by the
state of interest (hereafter referred to as the signal). Thus we
generalize the idea of a self-referencing detector, originally
introduced in 1977 by Klyshko \cite{Klyshko}, to different types
 of multi-photon states.  We use a time-multiplexed detector (TMD) to 
 emulate a photon-number-resolving detector.  The measured count statistics
$p(k)$ are related to the photon number distribution $\rho(n)$ by
$p(k)={\bf C} \cdot {\bf L}(\eta) \cdot \rho(n)$, where ${\bf
L}(\eta)$ is the matrix describing the binomial process of loss with
an overall efficiency of $\eta$ and ${\bf C}$ is a matrix that takes
into account that the TMD can only resolve up to a finite number of
photons at the input~\cite{LoopyJoint}\footnote{For a perfect
photon-number-resolving detector (even one with losses) this matrix
would be the identity.}. Thus the photon number distribution at the
source can be reconstructed from the count statistics by directly
inverting these matrices or by using a maximum likelihood technique.
We emphasize that we utilize this calibration technique to
accomplish, from a {\em single} measurement set, both loss
estimation and a reconstruction of the photon number distribution at
the source -- allowing loss-independent state characterization for a
given generation process without the need of a pre-measurement of
the loss.  As with Klyshko's original detector calibration scheme
our approach relies on knowledge that we infer from the state
generation; previous theoretical and experimental work indicates
that such inferences are reasonable \cite{Kumar, Mlynek,
YamamotoNON}. We emphasize that earlier
theoretical work \cite{Kiss95} has shown that for known losses such
compensation procedures are always possible when the overall
detection efficiencies exceed 50\% (and yet smaller efficiencies can
be tolerated for specific classes of states).

In general, any type of prior information can be used, but in this
Letter we use the strict photon number correlations between the
modes of a two-mode squeezed state; the state of the field is
\begin{equation}
\ket{\psi}=\sqrt{1-|\lambda|^2}\sum_{n=0}^{\infty} \lambda^n
\ket{n}_h\ket{n}_v, \label{eqn:squeezed}
\end{equation}
where the modes are labeled by orthogonal polarizations, $h$ and
$v$, and $\lambda$ is the parametric gain~\cite{Mandel}.  The state
we wish to characterize is the horizontally polarized mode (signal
arm) that is conditionally prepared by detection of the vertical
mode (trigger arm).  This would normally require an independent
measurement of the detector parameters and channel loss, but using
our technique can be inferred directly from the state
characterization data. Historically, detector calibration was
accomplished by assuming only the first two terms of the sum in
Eqn.~\ref{eqn:squeezed} contribute, measuring each mode with an APD
and comparing the singles counts to the coincidence counts. We
extend this idea in two ways: first, we incorporate $k$ terms of the
sum in Eqn.~\ref{eqn:squeezed}, where $k$ denotes the number of
photons detected as a trigger, allowing the use of higher parametric
gains and the characterization of a broader range of states; second,
we detect the signal with the TMD, allowing us to see the complete
count statistics of the signal, from which we then derive the
losses.


\begin{figure}
  \includegraphics[height=.19\textwidth]{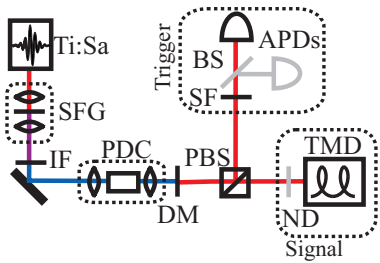}
  \caption{(Color online) The experimental setup: A mode-locked titanium sapphire laser
  (Ti:Sa) pumps sum frequency generation (SFG). The Ti:Sa is eliminated
  with Schott glass filters (not shown) and the SFG
  bandwidth is restricted with an interference filter (IF).  This is
  used to pump PDC in a waveguide and the blue is
  removed with a dichroic mirror (DM) and Schott glass filters (not shown).
  The PDC photons are split at a polarizing beamsplitter (PBS).  The
  trigger arm (reflected) contains a spectral filter (SF) and can be detected
  with either a single APD or a beamsplitter (BS) and two APDs
  detected in coincidence.  The signal arm (transmitted) is analyzed with the
  TMD.  The additional loss (ND) was optionally placed prior to the TMD.
    } \label{fig:setup}
\end{figure}


The experimental setup is shown in Fig.~\ref{fig:setup}.  A
mode-locked titanium sapphire (Ti:Sa) laser (100 fs pulses at 800 nm
and a repetition rate of 87 MHz) pumps second harmonic generation
(SHG), which we subsequently filter down to a 2 nm bandwidth (FWHM). Type
II PDC is generated in a 12 mm long z-cut KTP waveguide with 5
$\mu$W of second harmonic power. The orthogonally polarized daughter
photons split into different spatial modes at a polarizing
beamsplitter (PBS). The trigger arm is spectrally filtered with a 15
nm filter (FWHM). Details of this high brightness, waveguided
downconversion source are presented elsewhere~\cite{WavyPRL}. For
single photon generation, the trigger is a multimode fiber-coupled
avalanche photodiode (APD) (Perkin-Elmer SPCM-AQR-13).
Alternatively, for two photon state preparation, an additional BS is
inserted into the trigger arm and two APDs are detected in
coincidence ($3$ ns coincidence window) as a trigger.  A trigger is
only accepted if it occurs in a well-defined 700 ps window relative
to the Ti:Sa pump pulse. The signal arm is detected with the TMD.

\begin{figure}
  \includegraphics[height=.45\textwidth]{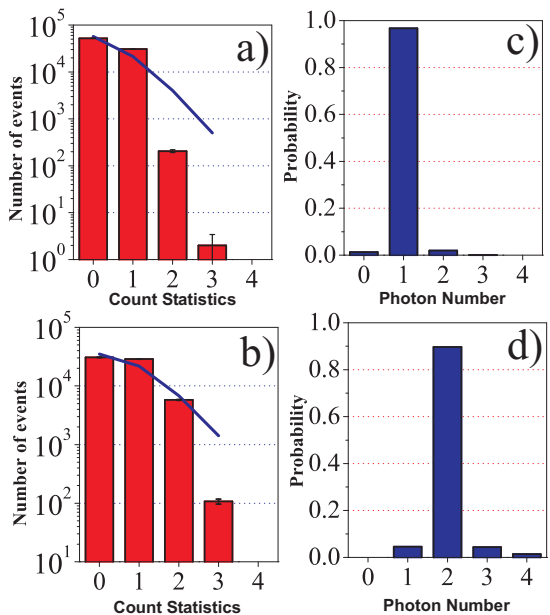}
  \caption{(Color online) Count statistics detected (logarithmic scale)
  using (a) single and (b) double APD as a trigger.  The solid line (used to guide the eye) shows
  the Poisson distribution with the same mean photon number as the data (0.376 and 0.623, respectively).  This line illuminates
  the sub-Poissonian nature of our measured statistics.  
  The photon
number distribution obtained by using a maximum likelihood inversion with constraints $\rho(n) \ge 0 $ for (c) single and (d) double APD trigger by taking
into account the efficiencies, which were $37.3\%$ and $31.5\%$, respectively.
  } \label{fig:data}
\end{figure}

Our results are presented in Fig.~\ref{fig:data}.  We begin with the
case of single APD trigger (Fig.~\ref{fig:data} a\&c).  Note that if
an APD was used in the signal arm this setup would be equivalent to
that of Klyshko. However, the use of the TMD allows us to verify the
complete count statistics of the conditionally prepared state that
can be used to verify the validity of of our assumption that
$\lambda$ is small such that higher order terms of
Eqn.~\ref{eqn:squeezed} are negligible.  We define the ``Klyshko
efficiency'' for our detector as\footnote{Klyshko defined the signal
efficiency in terms of detector click rates because he used a
continuous pump laser and therefore could not quantify $p(0|t=1)$.}
\[
\eta_K=p({\rm click}|{\rm click}) = \frac{\sum_{i=1}^{\infty} N_i}{
\sum_{i=0}^{\infty} N_i},
 \label{eqn:KlyshkoEff}
\]
where $N_i$ is the number events where $i$ photons would be
registered by a TMD and $p({\rm click|click})$ is the probability of
registering a click in the signal arm conditional on receiving a
click in the trigger arm. Using the TMD, we are able to define the
efficiency associated with single photon triggers as
\[
\eta=p(1|t=1) = \frac{N_1}{ \sum_{i=0}^{\infty} N_i},
 \label{eqn:OurEff}
\]
where $p(i|t=1)$ is the probability that $i$ photons were registered
in the signal arm given a single APD click in the trigger arm.  This
relation and the further relation, $p(0|t=1)=1-\eta$, can be used to
deduce an overall signal efficiency of $37.3\pm 0.1 \%$ from the
data.  Accounting for losses, this corresponds to a
single photon conditional preparation efficiency of $97\%$.

In our approach we utilize the complete conditional statistics to
ascertain $\eta$ from the experimentally independent measurements of
the different photon numbers $n$.  Note that in the general case of
$k$ photons detected in the trigger arm we expect $p(n<k|t=k)=0$ if
there is no loss or detector inefficiency ($\eta=1$), due to the
prior information of number correlations in the two modes.  Thus all
probabilities $p(n<k|t=k)>0$ are caused solely by losses in the
signal arm with
\begin{equation}
p(n<k|t=k)=\binom{k}{n}\eta^{n}(1-\eta)^{k-n}. \label{eqn:cond_loss}
\end{equation}
In this way we can exploit all such contributions with $n<k$ to
obtain a value for the efficiency independent of all other
experimental parameters like the parametric gain $\lambda$ and the
loss in the trigger arm.  We note that any state of light that has $\rho (n) =0$
for at least one value of $n$ (such as a single mode squeezed state where $\rho(n)=0$
for all odd values of $n$) is a perfect candidate for this technique.

For the double trigger (Fig.~\ref{fig:data}b\&d), one can calculate
the losses in the signal arm in three different ways using
Eqn.~\ref{eqn:cond_loss}. Given a two photon detection in the
trigger arm ($t=2$) and a signal efficiency $\eta$, it is
straightforward to calculate the following relations between these
probabilities and the efficiency in the signal arm:
%
\begin{eqnarray}
\eta^{(0)}&=&1-\sqrt{p(0|t=2)}
\\
\eta^{(1)}&=&{1 \over 2}[1-\sqrt{1-2p(1|t=2)} ]
\\
\eta^{(2)}&=&\sqrt{p(2|t=2)} \label{eqn:eff}
\end{eqnarray}
To evaluate the efficiencies $\eta^{(j)}$ from our raw data we must
consider not only the losses in our case, but also the limitations
of the photon-number-resolving capabilities of the TMD.  This is
done by multiplying our click statistics by the inverse of ${\bf C}$
and then applying Eqns.~3-5.  We
emphasize that this matrix can be obtained from the same measurement
as the characterization and no supplemental measurement is
necessary. Using the above relations we find that $\eta^{(0)}= 31.5
\pm 0.1 \%$, $\eta^{(1)}= 31.0 \pm 0.2\%$, and $\eta^{(2)}=32.1 \pm
0.2 \%$.  This corresponds to an average efficiency of $31.5\pm
0.2\%$ and a two-photon conditional state preparation efficiency of $90\%$.  The reason that these numbers differ by a degree larger than the error is because the state is not a true two-photon state, but contains small contributions from 
higher photon numbers. Qualitatively, it is simple to see that with higher photon-number contributions $\eta^{(2)}$ would be larger than its true value because $p(2|t=2)$ would be higher than if $\rho(n>2) = 0$. Simulations of our technique confirm this quantitatively. We note that this problem occurs in all calibration approaches using twin photon beams. However, in our case we are able to assess directly the validity of our assumptions by having access to the full count statistics, something that is impossible without photon-number resolution. Even with this consideration, our estimate of the signal arm efficiency (mean of above numbers with error bars encompassing the spread of values) is the most precise direct calibration available.
%

To test the accuracy of our loss estimation a neutral density filter
was inserted into the signal arm before the TMD, using the
double-APD-trigger configuration. The filter was calibrated to have
a transmission of $13.5\pm 0.1 \%$ using the Ti:Sa laser and a
linear photodiode. Using the previously mentioned loss relations a
measurement of the efficiency was performed with and without the
additional filter; The ratio of these two efficiencies gives a
filter transmission of
 $13.8\pm 0.1 \%$. The slight discrepancy in
transmittance is likely due to the spectral differences between the
laser and the PDC signal photons. (Non-degenerate PDC was used
resulting in a spectrum with a different central wavelength and
bandwidth from the laser. This calibration discrepancy reemphasizes
the need for a more accurate way to calibrate loss than with a prior
measurement using classical light.) As is expected for this case of very high
loss ($95.5\%$) the fidelity of the inverted distribution
decreases to 75\%.  Experimental attenuations performed with the single APD 
trigger were able to accurately reconstruct photon number distributions (showing $>90\%$ fidelity to a single photon state) for total signal loss of up to $88.0\%$.


An important case to study is when the prior information used (e.g.
photon number correlations) is not valid. The most extreme situation
would be the substitution of uncorrelated light sources for the PDC
source. We investigate this effect theoretically by mixing in a
coherent state into the PDC signal arm with various mean photon
numbers. We find that the fidelity of our photon number
reconstruction stays above 90\% for means up to $0.3$ photons,
independent of the loss; the additional mean number can be
significantly higher than 1 photon if the efficiencies are higher
than 55\% . Another key observation is that  decreasing prior
knowledge accuracy ({\em i.e.} increasing the mean photon number of
the additional coherent state) leads to inconsistent results among
the efficiency measurements. We confirm this experimentally for the
extreme case of completely uncorrelated light by pumping a waveguide
that is not periodically-poled and therefore produces no PDC but
creating spurious fluorescence counts. This yields count statistics
with a very low mean photon number, which we evaluate according to
our previous analysis. Using the false assumption of photon number
correlations results in: 1) inconsistent loss measurements and 2)
unphysical reconstructed photon number distributions (distributions
with negative probabilities)\footnote{Our measurement relies only on
photon-number correlations and is not compromised by impurity of the
state in Eqn~\ref{eqn:squeezed}.}.

Finally, to prove nonclassicality, we show that both our detected
statistics and the inferred photon number probabilities result in
negative values of Mandel's $Q$ parameter, which is a sign of nonclassicality. 
All Fock states result in $Q=-1$ and all coherent states have $Q=0$
The count statistics
detected using a single APD trigger give $Q=-0.36$ and the inferred
probability distribution result in $Q=-0.97$. Moreover, using two
APD triggering, the detected statistics yield $Q=-0.32$ and the
inverted distribution results in $Q=-0.93$.  To test the consistency
of this nonclassicality we also investigated the negativity of the P
function for our data. This measure can be formulated in terms of
conditions on the photon number distributions as~\cite{KlyshkoNeg,
Kumar}
\[
B(n)\equiv (n+2) \rho(n) \rho(n+2) - (n+1) [\rho(n+1)]^2 <0,
\label{eqn:B}
\]
where the inequality need only be satisfied for one value of $n$ to assure the
negativity of the P function.  Our detected statistics do not
satisfy the conditions, but our inferred photon probability
distributions do.  For the single [double] APD case, $B_1(0)=-0.13$
[$B_2(0)=-0.11$], once again showing the nonclassicality of the
states.

In summary, we have demonstrated a method of nonclassical state
characterization using mode-multiplexing and standard APDs.  An
inversion of the counting statistics was used in conjunction with a
`calibration-free' measurement of the loss in order to recreate the
photon number distribution of a heralded photon source based on the
two mode squeezed state. In the future, this technique could be
extended to more general nonclassical states where distinct
properties which change under the influence of loss are known. In
the context of conditional state preparation our characterization
technique can also be extended where loss calibration is not
obvious. By utilizing properties of known unconditioned statistics
we can estimate the losses in the system and thus obtain the
loss-tolerant calibration for post-selected conditioned subsets.
Hence we expect that our detection scheme will become particularly
relevant for quantum information protocols such as entanglement
distillation~\cite{BrowneED}.

This research was supported by the U.S. Department of Defense through the
Army Research Office grant DAAD19-02-1-0163.


%
%

\bibliographystyle{aipproc}   




\end{document}